\RequirePackage{xcolor}
\documentclass[journal,twoside,web]{ieeecolor}

\let\labelindent\relax
\usepackage{xcolor}
\usepackage{cite}
\usepackage{amsmath,amssymb,amsfonts}
\usepackage{algorithmic}
\usepackage{graphicx}
\usepackage{textcomp}
\usepackage{soul}
\usepackage{siunitx}
\usepackage[hidelinks]{hyperref}
\hypersetup{
  pdftitle={Multi-Sensor Edge Angle Detection for Performance Analysis in Ski Jumping},
  pdfauthor={Ivan Simeonov, Lukas Schulthess, Hanna Mueller, Marc Nölke, Michele Magno, Luca Benini, Christoph Leitner},
  pdfsubject={IEEE Sensors Journal},
  pdfkeywords={Assistive Technology, Feedback, Sports, Ultrasound, Time-of-Flight, Light-weight, Wearable}
}
\usepackage[nameinlink, capitalise]{cleveref}
\usepackage[nolist]{acronym}
\usepackage[flushleft]{threeparttable}
\usepackage{placeins}
\usepackage{svg}
\usepackage{orcidlink}
\usepackage[percent]{overpic}
\usepackage{mathtools}
\usepackage{multirow}
\usepackage{booktabs}
\usepackage{pbalance}
\usepackage{jsen}
\usepackage{wrapfig}
\usepackage{diagbox}
\usepackage{comment}
\usepackage{float}
\usepackage{tabularx}
\usepackage{makecell}
\usepackage{enumitem}

\usepackage{eso-pic}    % watermark

\newcommand{\lukas}[1]{}
\newcommand{\todo}[1]{}

\newcommand{\idea}[1]{}

\newcommand{\revise}[1]{#1}
\newcommand{\reviseSec}[1]{#1}

\newcommand{\RNum}[1]{\uppercase\expandafter{\romannumeral #1\relax}}
\DeclareRobustCommand{\IEEEauthorrefmark}[1]{\smash{\textsuperscript{\footnotesize #1}}}

\DeclareSIUnit\db{dB}
\DeclareSIUnit\dbi{dBi}
\DeclareSIUnit\dbm{dBm}
\DeclareSIUnit\watthour{Wh}
\DeclareSIUnit\mbps{Mbps}
\DeclareSIUnit\kbps{kbps}
\DeclareSIUnit\bps{bps}
\DeclareSIUnit\mAh{mAh}
\DeclareSIUnit\msInference{ms/inference}

\def\BibTeX{{\rm B\kern-.05em{\sc i\kern-.025em b}\kern-.08em
    T\kern-.1667em\lower.7ex\hbox{E}\kern-.125emX}}

\definecolor{abstractbg}{rgb}{0.89804,0.94510,0.83137}

\begin{acronym}

    \acro{BAN}{Body Area Network}
    \acro{PAN}{Personal Area Network}
    \acro{IoB}{Internet of Bodies}
    \acro{AI}{Artificial Intelligence}
    \acro{MCU}{Microcontroller}
    \acro{Nb-IoT}{Narrowband IoT}
    \acro{LoRa}{Long Range}
    \acro{UWB}{Ultra-Wideband}
    \acro{NFC}{Near Field Communication}
    \acro{VR}{Virtual Reality}
    \acro{AR}{Augmented Reality}
    \acro{EQS-HBC}{Electro-Quasistatic Human Body Communication}
    \acro{EQS}{Electro-Quasistatic}
    \acro{HBC}{Human Body Communication}
    \acro{WBAN}{Wireless Body Area Network}
    \acro{WPAN}{Wireless Personal Area Network}
    \acro{WP}{Work Package}
    \acro{CSMA}{Carrier Sense Multiple Access}
    \acro{KB}{Kilobyte}
    \acro{PMIC}{Power Management IC}

    \acro{HCI}{Human-Computer Interaction}
    \acro{HMI}{Human-Machine Interaction}
    \acro{TIA}{Transimpedance Amplifier}
    \acro{SoTA}{State-of-The-Art}
    \acro{WLAN}{Wireless Local Area Network}
    \acro{PCE}{Power Conversion Efficiency}
    \acro{ECG}{Electrocardiogram}
    \acro{BOM}{Bill of Material}
    \acro{DAQ}{Data Acquisition}
    \acro{RSSI}{Received Signal Strength Indicator}
    \acro{RSS}{Received Signal Strength}
    \acro{GBP}{Gain-Bandwidth Product}

    \acro{LoS}{Line-of-Sight}
    \acro{nLoS}{non-Line-of-Sight}
    \acro{QoS}{Quality of Service}
    \acro{NB}{Narrowband Communication}
    \acro{WPT}{Wireless Power Transfer}
    \acro{IBPT}{Intra-Body Power Transfer}
    \acro{ICNIRP}{International Commission on Non-Ionizing Radiation Protection} 
    \acro{SAR}{Specific Absorption Rate}

    \acro{RF}{Radio Frequency}
    \acro{SNR}{Signal-to-Noise Ratio}
    \acro{RFID}{Radio Frequency Identification}
    \acro{IoT}{Internet of Things}
    \acro{IoUT}{Internet of Underwater Things}
    \acro{UWN}{Underwater Wireless Network}
    \acro{UWSN}{Underwater Wireless Sensor Node}
    \acro{UAC}{Underwater Acoustic Channel}
    \acro{FSK}{Frequency Shift Keying}
    \acro{OOK}{On-Off Keying}
    \acro{ASK}{Amplitude Shift Keying}
    \acro{UUID}{Universal Unique Identifier}
    \acro{PZT}{Lead Zirconium Titanate}
    \acro{AC}{Alternating Current}
    \acro{NVC}{Negative Voltage Converter}
    \acro{NVCR}{Negative Voltage Converter Rectifier}
    \acro{FWR}{Full-Wave Rectifier}
    \acro{GPIO}{General Purpose Input/Output}
    \acro{PCB}{Printed Circuit Board}
    \acro{AUV}{Autonomous Underwater Vehicle}
    \acro{BLE}{Bluetooth Low Energy}
    \acro{FSR}{Force Sensing Resistor}
    \acro{SiP}{System in Package}
    \acro{SoC}{System on Chip}
    \acro{SpO2}{Oxygen Saturation}
    \acro{PULP}{Parallel Ultra-Low Power}
    \acro{ADC}{Analog to Digital Converter}
    \acro{TCDM}{Tightly Coupled Data Memory}
    \acro{GNSS}{Global Navigation Satellite System}
    \acro{IC}{Integrated Circuit}
    \acro{POM}{Polyoxymethylene}
    \acro{RTOS}{Real-Time Operating System}
    \acro{LoRaWAN}{Long Range Wide Area Network}
    \acro{ML}{Machine Learning}
    \acro{IIR}{Infinite Impulse Response}
    \acro{PLL}{Phase-locked Loop}
    \acro{RMS}{Root Mean Square}
    \acro{FPC}{Flexible Printed Circuit}
    \acro{I2C}{Inter Integrated Circuit}
    \acro{UART}{Universal Asynchronous Receiver-Transmitter}

    \acro{ToF}{Time of Flight}
    \acro{US}{Ultrasound}
    \acro{STR}{Static Target Rejection}
    \acro{RRR}{Reflection Rejection Range}

    \acro{FIR}{Finite Impulse Response}
    \acro{LIDAR}{Light Detection and Ranging}
    \acro{AoA}{Angle of Arrival}
    \acro{FoV}{Field of View}
    \acro{MAE}{Mean Absolute Error}
    \acro{IMU}{Inertial Measurement Unit}

    \acro{CFD}{Computational Fluid Dynamics}
    \acro{FIS}{Fédération Internationale de Ski}
    \acro{GRF}{Ground Reaction Forces}

    \acro{ECA}{Efficient Channel Attention}

\end{acronym}

\begin{document}

%=============================================
% Watermark for distribution
%=============================================
% Watermark for distribution ---------------------------------
\AddToShipoutPictureBG*{
  \AtPageUpperLeft{%
    \put(0,-40){\raisebox{23pt}{\makebox[\paperwidth]{\begin{minipage}{21cm}\centering
      \textcolor{gray}{This work has been accepted for publication in the IEEE Sensors Journal, \\
        DOI: 10.1109/JSEN.2026.3725626
      } 
    \end{minipage}}}}%
  }
  \AtPageLowerLeft{%
    \raisebox{20pt}{\makebox[\paperwidth]{\begin{minipage}{21cm}\centering
      \textcolor{gray}{\copyright 2026 IEEE. Personal use of this material is permitted. Permission from IEEE must be obtained for all other uses, in any current or future \\
        media, including reprinting/republishing this material for advertising or promotional purposes, creating new collective works, for resale or \\
        redistribution to servers or lists, or reuse of any copyrighted component of this work in other works.
        % https://doi.org/10.1145/3628353.3628549
      }
    \end{minipage}}}%
  }
}
%=============================================
% Main Section
%=============================================

\title{Multi-Sensor Edge Angle Detection for Performance Analysis in Ski Jumping}

\author{
    Ivan Simeonov\;\orcidlink{0009-0007-8840-5000}\IEEEauthorrefmark{1},
    Lukas Schulthess\;\orcidlink{0000-0002-6027-2927}\IEEEauthorrefmark{1} \IEEEmembership{(Graduate Student Member, IEEE)},\\
    Hanna Mueller\;\IEEEmembership{(Graduate Student Member, IEEE)},
    Marc Nölke,
    Michele Magno\;\orcidlink{0000-0003-0368-8923}\IEEEmembership{(Fellow, IEEE)},\\
    Luca Benini\;\orcidlink{0000-0001-8068-3806}\IEEEmembership{(Fellow, IEEE)},
    Christoph Leitner\;\orcidlink{0000-0002-7058-7236}\IEEEmembership{(Senior Member, IEEE)}
    \thanks{\IEEEauthorrefmark{1}Both authors contributed equally to the paper}
    \thanks{This work was supported in part by the CHIST-ERA project "SNOW" (Grant 209675) and the SNSF Ambizione Grant 233457.}
    \thanks{Ivan Simeonov, Lukas Schulthess, Hanna Mueller, Michele Magno, Luca Benini and Christoph Leitner are with the Department of Information Technology and Electrical Engineering, ETH Zürich, 8092 Zürich, Switzerland; Luca Benini is also with the Department of Electrical, Electronic, and Information Engineering, University of Bologna, Bologna, Italy; Marc Nölke is a certified trainer with the German Olympic Sports Confederation (DOSB) and a former competitive ski jumper.}
    \thanks{Corresponding emails: lukas.schulthess@pbl.ee.ethz.ch and christoph.leitner@iis.ee.ethz.ch}
    }

\IEEEtitleabstractindextext{%
    \fcolorbox{abstractbg}{abstractbg}{%
    \begin{minipage}{\textwidth}%

        \begin{wrapfigure}[20]{r}{3.5in}%
            \includegraphics[width=3.4in]{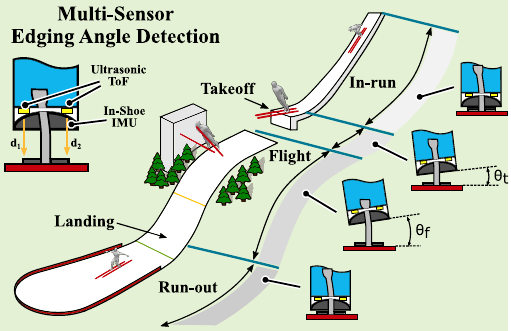}%
        \end{wrapfigure}%

\begin{abstract}
In ski jumping, performance during the gliding phase depends on achieving an aerodynamic posture that maximizes the lift-to-drag ratio.
In the V-style technique, the ski edge angle is a key determinant. Reducing the edge angle flattens the skis, increases their effective surface area, and improves aerodynamic lift, ultimately contributing to longer flight distances. Ski edge angles are biomechanically constrained by the limited range of ankle inversion. Increasing ankle inversion initially reduces the ski edge angles, thereby improving aerodynamic efficiency. However, inversion beyond the controllable biomechanical range may result in unstable edge configurations and asymmetric pressure distributions, thereby degrading aerodynamic efficiency and flight stability. Current sensing solutions widely quantify these angles using multi-system approaches that combine sensor signals through geometric relations. Such configurations require instrumentation on both the boot and the ski, altering mass distribution, affecting balance during flight, and increasing system complexity.
To overcome these limitations, this work presents a wearable sensing system that measures both boot inclination and ski edge angle without modifying the ski surface. Two ultrasonic \ac{ToF} sensors and an in-shoe \ac{IMU} are integrated into a single boot-mounted unit. Edge angles are estimated by combining ultrasonic distance measurements with \ac{IMU} data through geometric reconstruction of the boot–ski configuration.
Laboratory experiments demonstrate an angle resolution of \reviseSec{0.4500\textdegree}, a \ac{MAE} of \reviseSec{0.2640\textdegree}, and a coefficient of determination exceeding 99\% when compared with reference measurements, indicating strong linear agreement between the two modalities. The system achieves an end-to-end latency of 30.31\,ms, enabling real-time feedback suitable for athlete training, while consuming \revise{1.28\,mW} of power. With a total weight of only 18.6\,g the proposed system enables unobtrusive \reviseSec{measurement of ski edge angle} and boot orientation.
\end{abstract}

\end{minipage}}
}
\vspace{-10mm}

\maketitle

%\acresetall
\section{Introduction}\label{sec:intro}
\IEEEPARstart{S}{ki} jumping is one of the most spectacular Olympic sports and also one of the most technically demanding disciplines.
To maximize performance, ski jumpers must precisely control their posture and continuously and swiftly adapt to the demands of each phase of the jump \cite{j:Muller2009}. In a brief window of less than 10 seconds, athletes progress through four distinct stages in the jump: the in-run, take-off, flight, and landing \cite{c:Muller2005}.
While airborne, lift and drag forces must be carefully balanced to quickly achieve and sustain a stable glide \cite{elfmark_kinematic_2022}, hence maximizing jumping distance \cite{Schwameder01012008}. This aerial phase is decisive, as it \reviseSec{has} the longest duration and becomes increasingly critical with larger hills \cite{ELFMARK2022}.

Optimizing the body and ski posture during this phase is therefore essential to minimize drag and maximize lift.
The V-style technique in ski jumping, introduced in the 1980s, has been widely adopted due to its aerodynamic advantages over the previously used H-style \cite{Schwameder01012008}. In this technique, the ski tips are spread apart to form a "V" shape shortly after take-off. From a biophysical perspective, the V-style enables athletes to adopt a more forward-leaning posture, increasing the projected frontal surface area of the skis and body exposed to airflow. This maximizes aerodynamic benefits by improving the lift-to-drag ratio \cite{muller_performance_2008}, allowing jumpers to maximize jumping length while enhancing lateral stability \reviseSec{\cite{Watanabe2004}}.

Equipment, particularly suits and skis, significantly impacts the aerial phase of ski jumping \reviseSec{\cite{virmavirta_ski_2017}}.
The former is a predetermined parameter that can be favorably matched to an athlete's anthropometry within regulations of the \ac{FIS}. The latter can be actively utilized and controlled by the athlete. Through fine sensorimotor movements, the attitude (pitch, yaw, and roll) of the skis can be adjusted to the flight situation, directly influencing the lift-to-drag ratio \cite{muller_performance_2008}.
Several studies have investigated the ski attitudes in the
V-style technique with a focus on the angle-of-attack \reviseSec{\cite{Jung2014, Zhang2022}} (angle between the longitudinal axis of the body and the horizontal plane of the center of gravity), the ski-opening angle (two times yaw angle), and ski edge angle \reviseSec{\cite{virmavirta_aerodynamics_2019}} (roll angle).
Once a stable flight position is achieved, ski edge angles can be decisive in further increasing surface area \cite{virmavirta_ski_2017}, improving the lift-to-drag ratio, and ultimately enhancing flight distance.

\revise{A key biomechanical factor influencing the ski edge angle is the limited range of ankle inversion. The maximum achievable inversion during ski jumping is approximately \SI{20}{\degree}} \reviseSec{\cite{virmavirta_aerodynamic_ski_2019}}\revise{, which constrains the athlete’s ability to achieve a fully flat ski position in the V-style \cite{Cutter1994}. Ankle inversion is stabilized by the calcaneofibular and talofibular ligaments, as well as the peroneal muscles, which provide both static and dynamic lateral support. While these structures limit further reduction of the ski roll angle, they are essential for maintaining control against aerodynamic forces. To partially overcome this biomechanical constraint, equipment-based solutions have been explored. A notable example is the use of curved metal bindings, which gained attention at the 2010 Winter Olympics, where Switzerland's Simon Ammann achieved a significant performance advantage~\cite{FIS_Results_2010}. These bindings enable a reduction of the ski roll angle, allowing the skis to adopt a flatter orientation, thereby increasing effective surface area, improving the lift-to-drag ratio, and ultimately enhancing flight distance.}
However, excessive ankle inversion or overly curved bindings can lead to asymmetric loading of the skis, introducing unbalanced rolling moments that degrade flight stability and, consequently, flight distance \cite{virmavirta_aerodynamic_ski_2019, marquesbruna_mechanics_2009}.
\ac{CFD} simulations have shown that the optimal ski orientation has an angle of attack of \qty{30}{\degree}, a yaw angle of \qty{20}{\degree}, and a roll angle of \qty{0}{\degree}~\cite{Zhang2022}. \revise{This highlights the need for precise and objective measurement of the ski edge angle to help athletes identify and maintain the optimal configuration and maximize jump distance.}

Traditionally, performance assessment is performed via visual inspection \cite{j:sigrist2015}.
Jumps are recorded from the coaching tower and directly evaluated by trainers, who provide verbal feedback to improve the athletes' posture and dynamics \cite{schulthess_skiboot_2023}.
This approach relies heavily on the coach's experience, since no objective metrics can be directly extracted.

To reduce subjectivity and provide quantifiable metrics, computer vision techniques have been introduced, enabling more detailed analyses from video data.
Dunnhofer et al. \cite{dunnhofer_video_analysis_skijumping_2021} proposed a video-based approach for reconstructing ski jumping trajectories by using deep learning-based algorithms, including visual tracking and key-point detection to determine athletes' motion from an uncalibrated multi-camera system.
Similarly, Bao et al. \cite{bao_skijumping_pose_est_2023} presented a method based on monocular, high-resolution, and high-speed (\qty{500}{\hertz}) RGB camera recordings. They applied a HRNet with an \ac{ECA} extension (ECA-HRNet) to estimate human keypoints and track hip and knee angles during jumping, achieving an average precision of \qty{73.4}{\percent} on the COCO2017 test-dev set and the average precision of \qty{86.4}{\percent} on the KDSJ test set using the ground truth bounding boxes.

While vision-based systems can capture kinematic information, their performance is constrained by resolution and environmental factors.
Small changes, such as variations in the ski edge angle, are hard to detect from a distance.
This is a critical limitation, since the ski edge angles strongly influence aerodynamics and thereby affect flight stability and landing performance \cite{you_skijumping_simulation_2023}.
Moreover, their visibility can be further reduced by background interference or even completely disrupted by adverse weather conditions such as fog, snow, or rain \cite{Linhoff2022}.
In addition, accurate tracking often requires specific camera positions and viewing angles, which cannot be ensured throughout the entire jump.

To overcome these limitations, an alternative approach is to attach sensor systems directly to the athletes or their equipment \cite{brock_sensor_ski_jumping_2017}.
Small wearable \ac{IMU}-based platforms mounted at key motion points can capture kinematic data at the location of interest, providing detailed information on motions that are often visually obstructed \cite{hua_cps_ski_sports_2025, a_li_skijumping_tech_2024, Bessone2018}.
For example, Chardonnes et al. \cite{chardonnes_skijump_kinematics_2013} used wearable, \ac{IMU}-based systems to measure the orientation of the lower-body segments and skis during the entire jump sequence.
Logar et al. \cite{logar_wearable_2015} placed \ac{IMU}-based data loggers on each ski in front of the bindings, the athletes' lower legs, thighs, upper arms, and the sacrum to estimate the forces and moments in the ankle, knee, and hip of a ski jumping athlete during the in-run and take-off.
To optimize the landing phase, Bessone et al. \cite{bessone_skijunping_wearable_2019} used data collected by \acp{IMU} mounted to the skis, along with wireless force insoles, to analyze the correlation between ski movements and the \ac{GRF}.
More recently, Yu et al. \cite{yu_insole_skijumping_2023} merged \ac{IMU} and pressure insole data with 2D video recordings to investigate characteristics during phase transitions.
A comprehensive overview of these studies is given in \cref{tab:related_work_summary}.
All these approaches involve placing sensors directly on the skis, which may influence balance during flight, introduce disturbances, and conflict with \ac{FIS} regulations \cite{FIS2024}.
Furthermore, while additional sensor nodes can provide richer information, they also increase weight and may affect performance.

\begin{table*}[t]
\centering
\caption{Overview of studies that investigated ski roll attitude (edging angles) using wearable systems.}
\label{tab:related_work_summary}

\begin{threeparttable}
\begin{tabularx}{\linewidth}{
    >{\hsize=1.0\hsize\raggedright\arraybackslash}X  % 1.5× wider first column
    >{\hsize=1.3\hsize\centering\arraybackslash}X    % normal width
    >{\hsize=0.7\hsize\centering\arraybackslash}X    
    >{\hsize=1.2\hsize\centering\arraybackslash}X
    >{\hsize=1.0\hsize\centering\arraybackslash}X
    >{\hsize=1.0\hsize\centering\arraybackslash}X
    >{\hsize=1.0\hsize\centering\arraybackslash}X
    >{\hsize=0.7\hsize\centering\arraybackslash}X
}

\toprule
\textbf{Study}
 & \textbf{Sensor(s)}
 & \boldmath{$f_s^\dagger$}\;\textbf{[Hz]}
 & \textbf{Mounting}
 & \textbf{Output} 
 & \textbf{Device} 
 & \textbf{Size [mm]} 
 & \textbf{Weight [g]} \\
\midrule

\makecell[l]{Chardonnes et al.\\2013 \cite{chardonnes_skijump_kinematics_2013}} & \ac{IMU} & 500 & Behind binding & Ski attitude & Physilog & 50\,×\,20\,×\,15 & \textless\;100\\
\addlinespace[4pt]

\makecell[l]{Logar et al.\\2015 \cite{logar_wearable_2015}} & \ac{IMU} & 400 & Fore of binding & Ski attitude & Custom & 30\,×\,20\,×\,5 & \textless\;30 \\
\addlinespace[8pt]

\multirow{2}{*}[0pt]{\shortstack[l]{Bessone et al.\\2019 \cite{bessone_skijunping_wearable_2019}}}
 & \ac{IMU} & 100 & Behind binding & Ski attitude & MSR Solutions & \textless\;30\,×\,30\,×\,15 & \textless\;30 \\
 & Force Insoles & 100 & In boot & GRF & loadsol & \textless\;280\,×\,108\,×\,3.4 & \textless\;16 \\
\addlinespace[8pt]

\multirow{2}{*}[0pt]{\shortstack[l]{Yu et al.\\2023\cite{yu_insole_skijumping_2023}}}
 & \ac{IMU} & 400 & Fore of binding & Ski attitude & Xsens & \textless\;36\,×\,24\,×\,10 & \textless\;15 \\
 & Force Insole & 50 & In boot & GRF & Moticon & \textless\;290\,×\,105\,×\,4 & \textless\;15 \\
\addlinespace[8pt]

\midrule
\multirow{2}{*}[0pt]{\textbf{This work}} 
 & IMU & 100 & In boot & Boot attitude & Custom & 31\,×\,18\,×\,2.8 & 3.6 \\
 & Ultrasound & 50 & On boot & Boot-ski angle & Custom & 48\,×\,28\,×\,13 & 13.3 \\

\bottomrule
\end{tabularx}

\begin{tablenotes}
    \footnotesize
    \item[$^\dagger$] $f_s$ denotes the sampling frequency of the sensor.
\end{tablenotes}
\vspace{-5mm}

\end{threeparttable}
\end{table*}

To address these limitations, this paper presents a wearable sensing system that integrates boot inclination and ski edge angle measurement into a single boot-mounted unit without modifying the ski surface. Two ultrasonic \ac{ToF} sensors and an in-shoe \ac{IMU} are combined to estimate the ski edge angle via geometric reconstruction of the boot–ski configuration, eliminating the need for an additional ski-mounted sensor and reducing overall system weight, which is critical in ski jumping.

In particular, this article makes the following contributions:

\begin{enumerate}
\item \textbf{Investigation of Ultrasonic and \ac{IMU} Data Fusion} evaluates the effectiveness of fusing ultrasonic distance measurements with \ac{IMU}-based boot inclination to estimate the ski edge angle using a single boot-mounted device.
\item \textbf{A Full Multi-sensor System Integration} and performance evaluation on regulation-compliant ski jumping equipment under laboratory conditions.
\item \textbf{Validation Methodology} combines mathematical derivation, feasibility analysis, and system integration to precisely assess the proposed approach.
\end{enumerate}

The remainder of this article is organized as follows:
\cref{sec:mathematical_model} introduces the mathematical model and outlines the concept of using ultrasonic \ac{ToF} sensors together with an \ac{IMU} for ski edge angle detection.
\cref{sec:feasibility} describes the feasibility study conducted to determine the optimal lateral spacing between the two ultrasonic \ac{ToF} sensors.
\cref{sec:integration} details the integration of the evaluated system into the ski boot and the laboratory evaluation setup.
\cref{sec:results} presents and discusses the findings from both the feasibility study and the integrated system tests.
\cref{sec:limitations} analyzes the limitations of the proposed method and suggests potential solutions.
Finally, \cref{sec:conclusion} summarizes the key contributions and provides concluding remarks.

\section{Mathematical Model}\label{sec:mathematical_model}
A mathematical model is first derived to formally describe the principle of ultrasonic edge angle detection. The model enables analysis of the angular resolution, providing the basis for subsequent feasibility and experimental evaluations.

\subsection{Model Derivation}\label{ssec:math_model}
\begin{figure}
    \centering
    \includegraphics[width=\columnwidth]{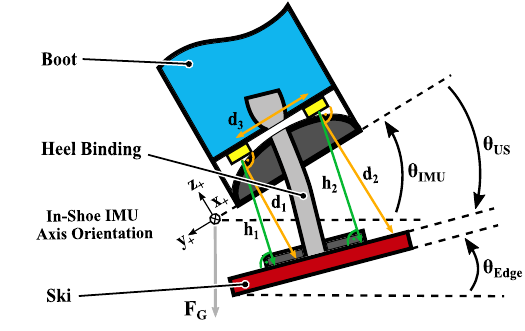}
    \vspace{-6mm}
    \caption{The 2D technical diagram showcases the principle of the edge angle acquisition $\theta_{Edge}$. It is directly related to $\theta_{US}$, derived from measured distances $h_1$ and $h_2$ and the inter-transducer $d_3$, and the angle between the soles plane and the gravity vector $\theta_{IMU}$.
    %The \ac{IMU} orientation shown above refers to a reference position in which the boot and ski lie flat on the ground, corresponding to $\theta_{\text{IMU}} = 0$ and $\theta_{\text{US}} = 0$.
    }
    \label{fig:mathematical_model}
\end{figure}

The core principle of our method for calculating ski edge angles is based on two angles: the counter-clockwise roll angle \revise{along} the x-axis \(\theta_{IMU}\) and the angle formed between the shoe and the ski's surface \(\theta_{US}\) (see \cref{fig:mathematical_model}).
From there, the ski edge angle \(\theta_{Edge}\) is defined as the angle between the ski's surface and the plane perpendicular to the gravity vector \(F_{G}\), which can be calculated:

\begin{equation}
    \theta_{Edge} = \theta_{IMU} - \theta_{US}
    \label{eq:0}
\end{equation}

\(\theta_{US}\) is determined using ultrasonic sensors that measure two distances, denoted as $h_1$ and $h_2$.
Each distance corresponds to the output of an ultrasonic \ac{ToF} sensor that measures the range to an acoustically reflecting plane, i.e., the ski. In the proposed setup, the sensors have a \ac{FoV} of \qty{45}{\degree}, ensuring that each sensor consistently returns the distance to the nearest point on the plane.
From these distance measurements, the perpendicular distances from the axis connecting the two transducers to the plane can be determined and are denoted as $d_1$ and $d_2$:

\begin{equation}
    d_1 = \frac{h_1}{\cos{\theta_{US}}},\quad d_2 = \frac{h_2}{\cos{\theta_{US}}}
    \label{eq:1}
\end{equation}

The distance $d_3$ between the two sensors is fixed, allowing the calculation of the angle $\theta_{US}$.
Using the trigonometric relationships of a right triangle

\begin{equation}
    \label{eq:2}
    \tan{\theta_{US}} = \frac{\vert d_1 - d_2 \vert}{d_3}
\end{equation}

Depending on the chosen reference orientation, the sign of the angle can be inverted by interchanging $d_1$ and $d_2$, due to the symmetry of the \reviseSec{$\arctan$} function. Substituting $d_1$ and $d_2$ from \cref{eq:1} into \cref{eq:2} yields

\[
    \tan{\theta_{US}} = \frac{(\frac{h_1}{\cos{\theta_{US}}} - \frac{h_2}{\cos{\theta_{US}}})}{d_3} = \frac{h_1-h_2} {d_3\cos{\theta_{US}}}
\]

which simplifies to:

\[
    \sin{\theta_{US}} = \frac{h_1-h_2}{d_3}
\]

Therefore, the final formulation for $\theta$, where $h_1$ and $h_2$ are the actual measured distances by the ultrasonic sensor, is:

\begin{equation}
    \theta_{US} = \arcsin\left(\frac{h_1 - h_2}{d_3}\right)
    \label{eq:4}
\end{equation}

This formulation remains valid as long as the critical angle condition is not violated: $\theta_{US} \le \tfrac{FoV}{2}$.

The second angle needed for \reviseSec{\cref{eq:0}} is calculated using the \ac{IMU} \revise{accelerations}:

\begin{equation}
    \theta_{IMU} = \arctan\left(\frac{-a_{y}}{\sqrt{a_{x}^2 +a_{z}^2}}\right)
    \label{eq:theta_IMU}
\end{equation}

\revise{where $\theta_{IMU}$ is the counter-clockwise roll angle around the x-axis, as illustrated in \cref{fig:mathematical_model}.}
%with \(a_{x}\) the acceleration along the x-axis and \(a_{z}\) the acceleration align the z-axis.

\subsection{Angular Resolution}\label{ssec:resolution_analysis}
To evaluate the feasibility of the proposed method, a theoretical analysis is conducted to determine the achievable angular resolution of $\theta_{US}$.
The resolution of the ultrasonic subsystem depends on the minimum detectable change in distance (\(\Delta h_\mathrm{min}\)) and the fixed spacing between the sensors \(d_{3}\). Defining the differential height between the two measured distances as

\[
    \Delta h = h_1 - h_2,
\]

The expression for the edge angle can be rewritten as:

\begin{equation}
    \label{eq:simplfied_theta}
    \theta_{US} = \arcsin\left(\frac{\Delta h}{d_3}\right).
\end{equation}

Based on \cref{eq:simplfied_theta}, the influence of the individual distance measurements on the angular resolution can be analyzed.
Define the uncertainties of \(h_1\) and \(h_2\) as \(\delta h_1\) and \(\delta h_2\), respectively. Since identical ultrasonic sensors are used, both uncertainties can be assumed to be equal. Furthermore, the two measurements are considered independent, since their noise sources are independent. Under these assumptions, the combined uncertainty in distance difference \(\Delta h\) is given by:

\[
    \delta (\Delta h) = \sqrt{(\delta h_1)^2 + (\delta h_2)^2} = \sqrt{2}\,\delta h.
\]

Using uncertainty propagation for \cref{eq:simplfied_theta}, the resulting uncertainty in $\theta_{US}$ is:

\begin{equation}
    \label{eq:derivative}
    \delta \theta = \left| \frac{\partial \theta_{US}}{\partial (\Delta h)} \right|\, \delta (\Delta h),
\end{equation}

From \cref{eq:derivative} we see that \(\delta \theta\) scales linearly with the measurement uncertainty of \(\delta h\). Therefore, maximizing the resolution of the distance measurements \(h_1\) and \(h_2\) directly enhances the precision of the calculated angle \(\theta_{US}\). This implies that for a given \(\theta_{US} \neq 0\):

\[
    \sin\theta_{US} = \frac{\Delta h}{d_3} \quad \Longrightarrow \quad \Delta h = d_3 \sin{\theta_{US}}
\]

Thus, increasing the inter-transducer distance \(d_3\) results in a proportionally larger \(\Delta h\). The derivative of \(\theta_{US}\) with respect to \(\Delta h\) is obtained as:

\begin{align*}
    \frac{\partial \theta_{US}}{\partial (\Delta h)} &= \frac{1}{d_3 \sqrt{ 1 - \left(\frac{\Delta h}{d_3}\right)^2} }\\[6pt]
                                 &= \frac{1}{d_3 \sqrt{1-\sin^2\theta_{US}}}\\[6pt]
                                 &= \frac{1}{d_3 \cos{\theta_{US}}}
\end{align*}

Inserting this derivative in the uncertainty propagation for \(\theta_{US}\) yields:

\begin{equation}
    \label{eq:computed_derivative}
    \delta \theta_{US} = \frac{1}{d_3 \cos{\theta_{US}}}\, \delta (\Delta h)
\end{equation}

\begin{figure}
    \centering
     \includegraphics[width=\columnwidth]{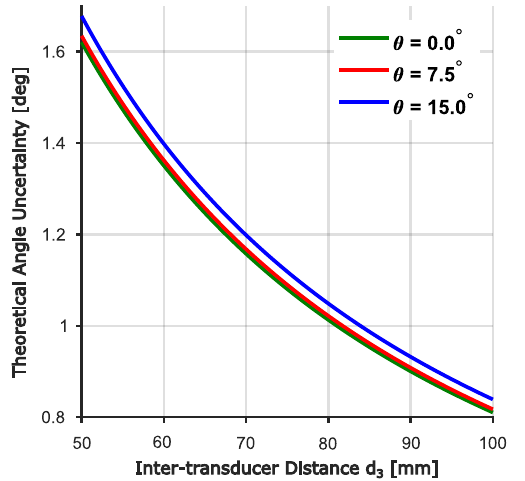}
    \vspace{-6mm}
    \caption{Theoretical angle uncertainty for different inclination angles \(\theta_{US}\) and variable inter-transducer distances.}
    \label{fig:lim}
\end{figure}

Therefore, a larger \(d_3\) produces a greater measurable difference between \(h_1\) and \(h_2\) for a fixed inclination angle \(\theta_{US}\), while simultaneously reducing the uncertainty propagation factor \(\tfrac{1}{d_3 \cos{\theta_{US}}}\).
\cref{fig:lim} presents the modeled uncertainty behavior according to the derived relationship. Overall, this analysis indicates that ultrasonic sensors with minimal uncertainty (i.e., higher ultrasonic frequencies) should be used and their inter-sensor distance \(d_3\) should be maximized.
Moreover, the theoretical uncertainty decreases with smaller inclination angles $\theta_{US}$.
With a theoretical angular uncertainty of less than \reviseSec{\ang{0.8500}}, the system provides sufficient resolution to capture the binding angle, which is typically limited to a maximum curvature of \ang{15}~\cite{virmavirta_aerodynamic_ski_2019}.

\section{Feasibility Study} \label{sec:feasibility}
With the mathematical foundation established, this section investigates the practical feasibility of the proposed edge angle detection method.

\begin{figure}
    \centering
    \begin{overpic}[width=\columnwidth]{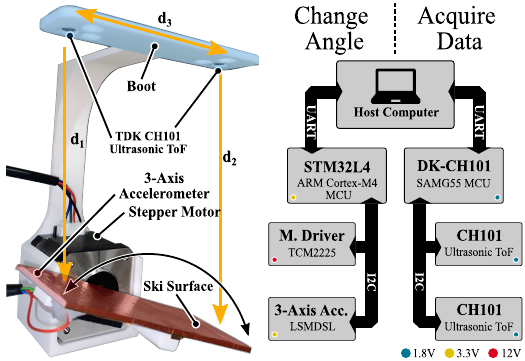}
        \put(0,65){(a)}
        \put(48,65){(b)}
    \end{overpic}
    \vspace{-5mm}
    \caption{Overview of the experimental setup: (a) Implementation of the evaluation setup, (b) high-level block diagram of the system control used for the feasibility evaluation.}
    \label{fig:eval_setup}
\end{figure}

\subsection{Testbench}\label{sec:testbench}
The study focuses on identifying the optimal sensor configuration and inter-sensor distance to validate the theoretical model under controlled laboratory conditions before integration into the ski boot. For this, a simplified yet versatile experimental setup shown in \cref{fig:eval_setup} (a) was designed.
\reviseSec{It consists of a 3D-printed test bench with a T-shaped head featuring mounting holes for three inter-sensor distances (\(d_3 = 60, 70, 80\,\mathrm{mm}\)). A \textit{NEMA 17} stepper motor at the base drives a \qty{110}{\milli\meter}-wide acrylic plate representing the ski cross-section. The plate is directly coupled (1:1) to the motor shaft, so each microstep maps to an identical plate rotation, making the plate's angular resolution equal to the motor's.}
This configuration allows systematic evaluation of different combinations of \(\theta_{US}\) and inter-sensor distances \(d_3\) under controlled conditions.

As ultrasonic sensors, the \textit{TDK InvenSense CH101} \ac{ToF} modules were selected for their low \ac{RMS} range noise of \qty{1}{\milli\meter}, which reduces the achievable angular uncertainty as discussed in \cref{eq:computed_derivative}. \reviseSec{Although optical reflective \ac{ToF} sensors were considered, their performance is highly sensitive to environmental conditions such as ambient sunlight. For example, the \textit{ST VL53L4CD} exhibits a reduction in ranging capability accuracy to $\pm$\,\qty{9}{\milli\meter} under elevated ambient light, which corresponds to 6.5\,\% error in the short-range mode. Moreover, optical \ac{ToF} sensors generally consume significantly more power than ultrasonic alternatives, which is critical for battery-operated systems. The \textit{ST VL6180X} consumes \qty{61}{\milli\watt}, compared to only \qty{34}{\micro\watt} for the \textit{TDK InvenSense CH101}.} Furthermore, the ability of the \textit{CH101} to measure short distances down to \qty{9}{\centi\meter}, combined with a lightweight design of only \qty{0.1}{\gram} and a compact \revise{\(8\times8\times8\) \unit{\milli\meter}} package, enables robust integration with minimal impact on athlete comfort and aerodynamics. In addition, the ultrasonic sensors incorporate in-sensor data processing features such as \ac{STR} and \ac{RRR}. The \ac{STR} suppresses reflections from static objects within the measurement range, while the \ac{RRR} rejects echoes originating from distances shorter than a defined threshold. Together, these filtering mechanisms play a crucial role in obtaining representative distance readings by excluding unwanted reflections from static mechanical components, such as the ski binding.
During evaluation, the ultrasonic sensors measuring $h_1$ and $h_2$ were operated over a measurement range of \qty{15}{\centi\meter}, corresponding to the maximum binding extension.
To mitigate reflections from the nearby binding rod, the \ac{STR} and \ac{RRR} filters were set to \qty{4.5}{\centi\meter}, reflecting the expected proximity conditions after system integration.
Both transducers were positioned \qty{120}{\milli\meter} above a horizontally oriented reflection plane. At \(\theta\) = \ang{0}, this configuration results in equal distances of \revise{\(d_1 = d_2 =\) \qty{120}{\milli\meter}}, representing the maximum separation that occurs when the binding is fully extended.

\subsection{Data Recording and Evaluation}\label{sec:testbench_validation}
Based on the mathematical model derived in \cref{sec:mathematical_model}, the testbench was used to determine the system's maximum angle resolution. To conduct the experiments, a host computer controlled both the high-level angle adjustment and data acquisition over two independent subsystems communicating over a \ac{UART} interface (see \cref{fig:eval_setup} (b)). The ultrasonic sensors were configured and operated using a \textit{TDK InvenSense DK-CH101} development board. The testbench itself was controlled by a \textit{STMicroelectronics STM32L452RE} board interfaced with an \textit{Analog Devices TMC2208} motor driver, providing a step resolution of \ang{0.1125}. An \textit{STMicroelectronics LSM6DSV16BX} \ac{IMU} with an internal low-pass filter was mounted on the underside of the plate. The \ac{IMU} output served as a feedback for the stepper motor control and as ground truth for the measured inclination angles.

A series of measurements was conducted at different inclination angles of the reflection surface, using three theoretical angular step sizes of \reviseSec{\ang{0.5625}, \ang{0.4500}, and \ang{0.3375}} and three inter-sensor distances \(d_3\) of \qty{60}{\milli\meter}, \qty{70}{\milli\meter}, and \qty{80}{\milli\meter}. \reviseSec{These angular step sizes are multiples of the minimal angular step of \ang{0.1125} that can be achieved with the \textit{NEMA 17} and the \textit{TMC2208} motor driver.} For each configuration, 100 samples were recorded from both ultrasonic sensors and the \ac{IMU}. The collected data was used to calculate \(\theta_{US}\) according to \cref{eq:simplfied_theta}.

To assess the relationship between \(\theta_{US}\) and \(\theta_{IMU}\), the Pearson correlation coefficient (\(R^2\)) and \ac{MAE} were computed.
Statistical significance of the correlation was evaluated using a two-tailed t-test with a significance level of \(p = 0.001\).
The null hypothesis (\(H_0\)) assumes no linear relationship (\(R^2 = 0\)), whereas the alternative hypothesis (\(H_1\)) states a significant linear relationship (\(R^2 \neq 0\)).
To assess the agreement between the two measurement methods, a Bland–Altman plot was generated to visualize the mean difference and the limits of agreement between \(\theta_{US}\) and \(\theta_{IMU}\).

\begin{figure}[thb]
    \centering
        \begin{overpic}[width=\columnwidth]{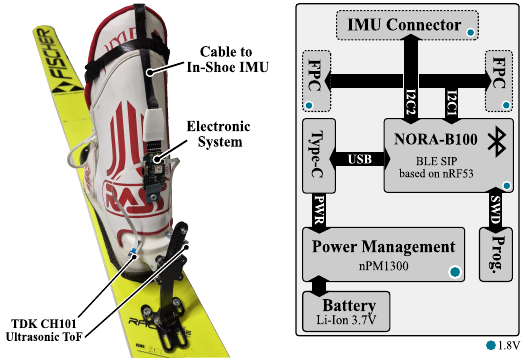}
        \put(2,66){(a)}
        \put(50,66){(b)}
    \end{overpic}
    \vspace{-5mm}
    \caption{Overview of the proposed system integration: (a) System integration onto the ski boot, (b) High-level block diagram of the electronic system.}
    \vspace{-6mm}
    \label{fig:on_shoe}
\end{figure}

\section{System Integration}\label{sec:integration}
Based on the feasibility study results, we designed and implemented a custom sensor node that combines ultrasonic distance sensing with an inertial measurement unit (IMU) to enable single-device ski edge angle estimation.

\subsection{Electronic Subsystem}\label{sec:electronic_subsystem}
The electronic subsystem serves as the central unit for data collection and processing. \cref{fig:on_shoe} (b) shows a simplified block diagram of the electronics.

\begin{figure*}[thb]
    \centering
    \includegraphics[width=\textwidth]{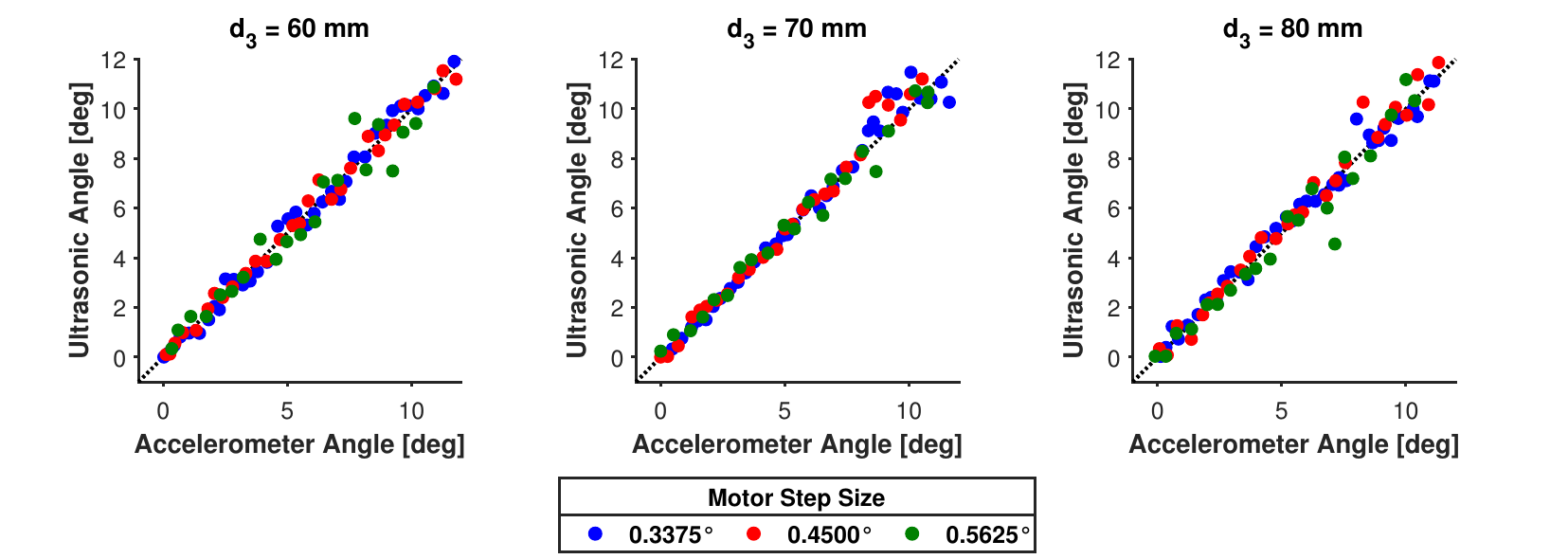}
    \vspace{-7mm}
    \caption{\reviseSec{Scatter plots comparing the accelerometer angle \(\theta_{IMU}\) and the ultrasonic angle \(\theta_{US}\) obtained with the proposed approach. Results are shown for three inter-transducer distances \(d_3 = 60, 70, 80\,\mathrm{mm}\) and three motor step size resolutions ($0.3375^\circ$ in blue, $0.4500^\circ$ in red, and $0.5625^\circ$ in green).}}
    \label{fig:imu_us_scatter}
\end{figure*}

\begin{figure*}[thb]
    \centering
    \includegraphics[width=\textwidth]{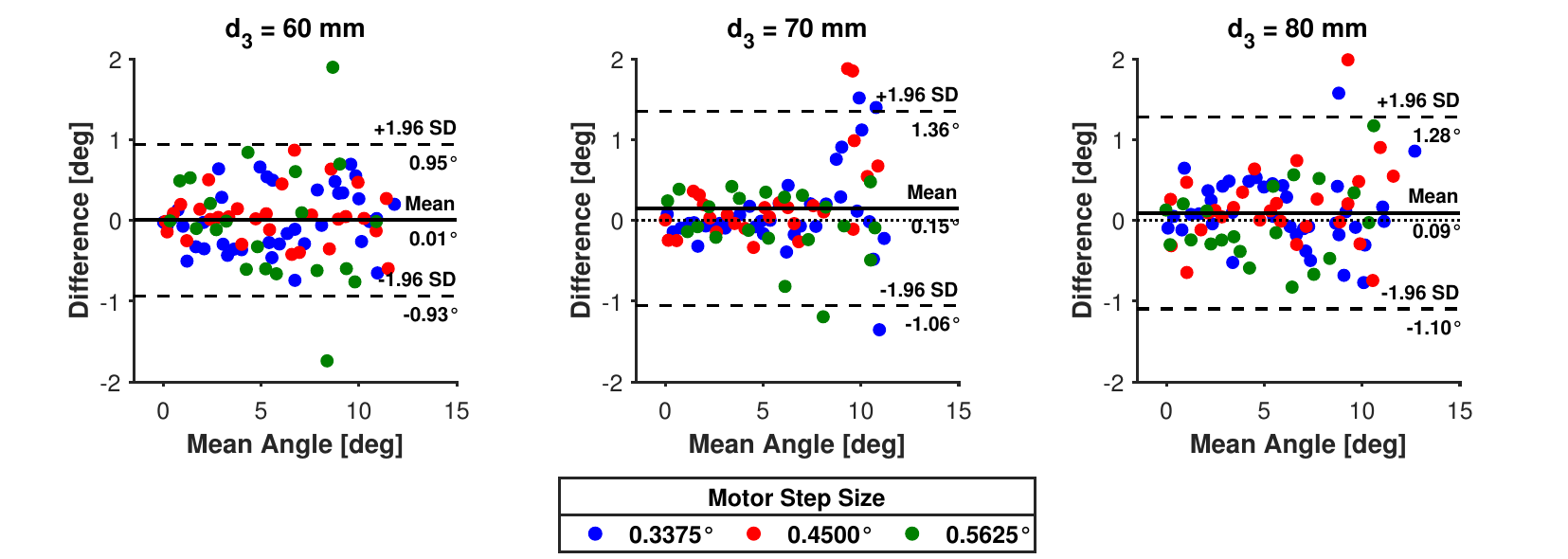}
    \vspace{-6mm}
    \caption{\reviseSec{Bland--Altman plots comparing the accelerometer angle \(\theta_{IMU}\) and the ultrasonic angle \(\theta_{US}\) obtained with the proposed approach. Results are shown for three inter-transducer distances \(d_3 = 60, 70, 80\,\mathrm{mm}\) and three motor step size resolutions ($0.3375^\circ$ in blue, $0.4500^\circ$ in red, and $0.5625^\circ$ in green). The horizontal axis represents the mean angle, and the vertical axis shows the difference between methods.}}
    \label{fig:bland_altman}
\end{figure*}

The system uses a \textit{Nordic Semiconductor nRF5340} integrated within a \textit{u-blox NORA-B100}. It manages communication with two \textit{TDK CH101} ultrasonic \ac{ToF} sensors via an \ac{I2C} interface, which are connected to the \ac{PCB} through \ac{FPC} cables.
A second \ac{I2C} bus connects a \textit{TDK ICM-20948} 9-axis \ac{IMU} positioned inside the shoe, lying flat within the plane of the sole. Power management is handled by the \textit{Nordic Semiconductor nPM1300} \ac{PMIC}, which monitors the state of charge, controls the charging of the \textit{Renata ICP341018PM} \qty{33}{\mAh} lithium-polymer battery, and generates the system-wide supply voltage of \qty{1.8}{\volt}. During laboratory testing, data are transmitted via USB to a host computer.

\subsection{Mechanical Integration}\label{sec:mechanical_integration}
\cref{fig:on_shoe}(a) illustrates the mechanical integration of the sensors and electronics. Direct placement of the \textit{TDK CH101} ultrasonic sensors on the cover of the shoe was not an option, as at higher angles, the ski boot was blocking the ultrasonic echoes reflected from the skis. That is why the sensors were placed farther from the shoe, using two \SI{1.5}{\centi\meter}-long 3D-printed mounts, spaced exactly \SI{70}{\milli\meter} apart. This allows for the precise adjustment of the position on the go without obstructing the pressure waves. On top of that, this approach does not interfere with detaching the bindings from the shoe. The electronic module is centrally mounted on the back of the shoe, representing a potential integration step for future system implementation. An extension cable provides flexibility in positioning the \ac{IMU} within the shoe.

\subsection{Data Recording and Evaluation}\label{sec:integration_validation}
To validate and test the integrated system, the ski boot was mounted in the ski binding, and the ski was placed on flat ground and stabilized using two \qty{5}{\kilo\gram} weights positioned in front of and behind the binding. An additional \qty{5}{\kilo\gram} weight was placed inside the boot to apply down-force during the experiment, emulating the load of an athlete. For repeatable measurements, a metal cord was mounted directly onto the binding and connected to a pulley system driven by a stepper motor. A \textit{Raspberry Pi} controlled the motor and enabled precise linear actuation, pulling the cord by a defined constant length to lift the boot and generate a specific and reproducible movement pattern.

Signal processing and computations were performed directly on the \ac{MCU} using firmware implemented with the Zephyr \ac{RTOS} to estimate the system's latency and power consumption. The \textit{TDK CH101} sensors are sampled at \qty{100}{\hertz}. The firmware uses each pair of distances to compute the ultrasonic angle $\theta_{US}$ as defined in \cref{eq:simplfied_theta}. The computed angle samples are passed through a 7-tap low-pass \ac{FIR} filter with a cutoff frequency of \qty{6}{\hertz} and a block size of 1 sample.
\reviseSec{The filter order was selected by optimizing a cost function that equally weighted \ac{SNR} and group delay for low-pass filters with 4 to 20 taps.
A filter with seven taps provided the best trade-off between noise reduction and latency.
The cutoff frequency was set to \qty{6}{\hertz} based on ski jumping dynamics. Since the transition from take-off to steady flight takes about \qty{0.7}{\second} \cite{braghin_skijumping_aerodynamics_2016,j:skijumping_takeoff_vodicar_2010}, this cutoff preserves the relevant motion dynamics while attenuating higher-frequency noise from sensors, vibrations, and transient effects.}

In parallel, the \textit{TDK ICM-20948} 9-axis \ac{IMU} attached to the ski boot sole is sampled at \SI{100}{\hertz}, using the same acquisition timing as the ultrasound sensors, providing one sample point for each computed ultrasonic angle $\theta_{US}$.
These \ac{IMU} measurements are used to compute the angle $\theta_{IMU}$ according to \cref{eq:theta_IMU}.
They are subsequently filtered with the same low-pass filter used for the ultrasonic sensor data.
Finally, the edge angle $\theta_{Edge}$ is obtained by subtracting the two angles, as shown in \cref{eq:0}.
This last operation is not evaluated in this paper, as we use the \ac{IMU} to assess the accuracy of the ultrasonic sensing modality.

\section{Results and Discussion}\label{sec:results}
This section presents the results of the progressive evaluation of the proposed method.
Although the primary focus of this work is to evaluate the feasibility of using ultrasonic sensors to measure the ski edge angle, this section provides estimates of key parameters, including device power consumption and latency characteristics, based on a custom-designed hardware platform.

\subsection{Feasibility Study}
The baseline evaluation was performed using the setup shown in \cref{fig:eval_setup}.
Three inter-sensor distances \revise{(\(d_3 = \) \qty{60}{\milli\meter}, \qty{70}{\milli\meter}, \qty{80}{\milli\meter})} and three theoretical angular step sizes (\ang{0.5625}, \revise{\ang{0.4500}}, \ang{0.3375}) were tested, resulting in nine configurations in total.
For each inter-sensor distance, all three step sizes were combined into a single correlation plot comparing the ultrasonic estimates \(\theta_{US}\) with the reference angles \(\theta_{IMU}\) obtained from the \ac{IMU} ground-truth data (\cref{fig:imu_us_scatter}).
All measurements covered an angular range of \SIrange{0}{11}{\degree} with 100 samples per step. Bland–Altman plots derived from the same data are presented in \cref{fig:bland_altman}.

\begin{table}[thb]
    \centering
    \renewcommand{\arraystretch}{1.2}
    \caption{$R^2$ coefficient of angle resolutions with respect to the inter-sensor distance $d_3$.}
    \label{tab:R2}
    \begin{tabularx}{\columnwidth}{@{}l>{\centering\arraybackslash}X>{\centering\arraybackslash}X>{\centering\arraybackslash}Xc@{}}
        \toprule
        & \multicolumn{3}{c}{\textbf{$\mathbf{R^2}$ Coefficient per Angular Step Size [deg]}} & \\
        \cmidrule(lr){2-4}
        \textbf{\textbf{$d_3$ [\si{\milli\meter}]}} & \textbf{0.3375} & \textbf{0.4500} & \textbf{0.5625} & \textbf{Mean} \\
        \midrule
        60 & 0.994 & 0.996 & 0.974 & 0.988 \\
        70 & 0.990 & 0.991 & 0.977 & 0.986 \\
        80 & 0.989 & 0.990 & 0.977 & 0.985 \\
        \midrule
        \textbf{Mean} & 0.991 & 0.992 & 0.976 & \\
        \bottomrule
    \end{tabularx}
\end{table}

\begin{table}
    \centering
    \renewcommand{\arraystretch}{1.2}
    \caption{Mean Absolute Error (MAE) of angle resolutions with respect to the inter-sensor distance $d_3$.}
    \label{tab:MAE}
    \begin{tabularx}{\columnwidth}{@{}l>{\centering\arraybackslash}X>{\centering\arraybackslash}X>{\centering\arraybackslash}Xc@{}}
        \toprule
        & \multicolumn{3}{c}{\textbf{MAE per Angular Step Size [deg]}} & \\
        \cmidrule(lr){2-4}
        \textbf{\textbf{$d_3$ [\si{\milli\meter}]}} & \textbf{0.3375} & \textbf{0.4500} & \textbf{0.5625} & \textbf{Mean} \\
        \midrule
        60 & 0.319 & 0.264 & 0.725 & 0.436 \\
        70 & 0.591 & 0.414 & 0.447 & 0.484 \\
        80 & 0.431 & 0.385 & 0.522 & 0.446 \\
        \midrule
        \textbf{Mean} & 0.447 & 0.354 & 0.565 & \\
        \bottomrule
    \end{tabularx}
\end{table}

\cref{tab:R2} and \cref{tab:MAE} summarize the performance metrics, namely the Pearson correlation coefficient (\(R^2\)) and mean absolute error (\ac{MAE}). Each table column corresponds to a tested step size, and each row to an inter-sensor distance. All configurations achieved statistically significant linear correlations (two-tailed t-test, \(p < 0.001\)).
The strongest agreement between \(\theta_{US}\) and \(\theta_{IMU}\) was observed for \(d_3 = \) \qty{60}{\milli\meter} and a step size of \revise{\ang{0.4500}}, with \(R^2 = 0.996\) and \reviseSec{\ac{MAE} = \ang{0.2640}}.

Although theoretical analysis in \cref{ssec:resolution_analysis} suggests that smaller \(d_3\) should increase measurement uncertainty, no consistent degradation in performance was observed across the tested distances. This results from the relatively small variations (\qty{10}{\milli\meter}) in \(d_3\) within the current setup, indicating that such scale changes have a negligible effect on system performance. Considering the angular step size, the smallest \ac{MAE} and highest correlation were obtained near \reviseSec{\ang{0.4500}}. For step sizes below \reviseSec{\ang{0.4500}}, performance declined, indicating the operational limit of the proposed sensing approach.

In both the scatter and Bland–Altman plots, a slight increase in variance is visible at higher inclination angles. This behavior aligns with the theoretical prediction in \cref{ssec:resolution_analysis}, where the measurement uncertainty increases nonlinearly with \(\theta\). At higher inclinations, the ultrasonic beam intersects the reflective surface at oblique angles, causing the \ac{ToF} sensor to register reflections from the nearest point within its \ac{FoV}. This geometric effect leads to larger deviations in the estimated angle.
The increase in variance is particularly evident for \revise{$d_3 = \qty{70}{\milli\meter}$} and \qty{80}{\milli\meter}, which can be attributed to the finite width of the reflective surface \reviseSec{(\qty{110}{\milli\meter})}. At large \(\theta\), the projected beam altitudes (\(h_1, h_2\)) extend toward the edge of the surface, violating the model assumption of an infinitely wide plane.
For \revise{$d_3 = \qty{60}{\milli\meter}$}, this effect remains negligible as the beams stay within the reflection boundary.

In summary, the results confirm a strong linear relationship between $\theta_{US}$ and $\theta_{IMU}$, with an effective angular resolution near \reviseSec{\qty{0.4500}{\degree}}.
Inter-sensor distances up to \qty{70}{\milli\meter} yield comparable performance, indicating that compact sensor integration is feasible without notable loss in accuracy.

\subsection{System Integration}
In this section, the performance of the proposed method is evaluated by integrating the system into a ski-jumping boot and testing it under controlled laboratory conditions. A direct comparison with the in-shoe \textit{TDK ICM-20948} \ac{IMU}, which serves as the reference system, is performed. This enables the computation of the \ac{MAE}, standard deviation, and latency. All presented results are corrected for latency to ensure an objective error analysis.

\begin{figure}
    \centering
    \includegraphics[width=\columnwidth]{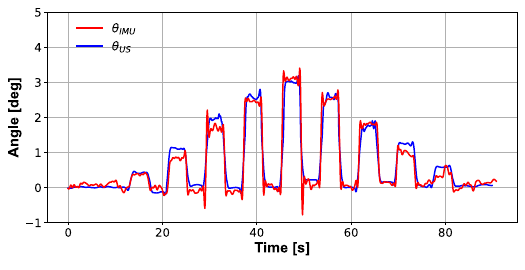}
    \vspace{-7mm}
    \caption{Comparison of $\theta_{IMU}$ and $\theta_{US}$ time during motion cycles.}
    \vspace{-4mm}
    \label{fig:us_step_response}
\end{figure}

The first performance test consisted of a step-response evaluation to assess the system's angle-tracking performance. The pulley executed five consecutive steps, starting at \qty{0}{\degree} and increasing the angle by \reviseSec{\qty{0.6000}{\degree}} per step up to \qty{3}{\degree}. For each step, both the elevated and lowered positions were held for \qty{5}{\second} and compared against the angle obtained from the \ac{IMU} data.
For this experiment, the raw \ac{IMU} data were filtered in post-processing using the same 7-tap low-pass filter used in the firmware, implemented as a zero-phase filter to preserve the phase and ensure an objective latency comparison.
The results of this experiment are shown in \cref{fig:us_step_response}.
From the graphs, it can be seen that $\theta_{US}$ is less sensitive to oscillations caused by the ski boot's rapid acceleration than $\theta_{IMU}$.
\revise{This can be attributed to the fact that the ultrasonic sensor inherently averages over both time and the spatial extent of the acoustic beam, effectively filtering out high-frequency vibrations that are directly captured by the \ac{IMU}.}
Furthermore, no significant hysteresis is observed, indicating that the system reliably detects repeated upward and downward movements.

\begin{figure}
    \centering
    \includegraphics[width=\columnwidth]{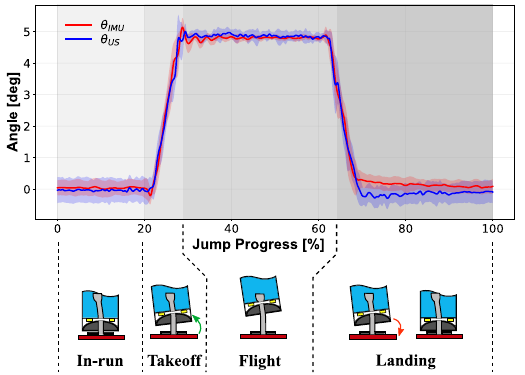}
    \vspace{-6mm}
    \caption{Temporal comparison of $\theta_{IMU}$ and $\theta_{US}$ with their confidence bands of 1-$\sigma$ throughout 10 simulated ski jumps, highlighting the four phases: in-run, take-off, in-flight, and landing.}
    \vspace{-6mm}
    \label{fig:simulated_jump}
\end{figure}

A second experiment focused on replicating a complete ski-jumping jump, consisting of the in-run, take-off, flight, and landing phases, which are highlighted in different colors in \cref{fig:simulated_jump}.
The y-axis shows the mean and standard deviation of $\theta_{US}$ and $\theta_{IMU}$ computed from 10 individual cycles. The x-axis represents the normalized jump phase, with corresponding images of the ski-boot position shown \revise{on the bottom of \cref{fig:simulated_jump}} for the reader's reference.
At the end of the takeoff, the \ac{IMU} shows damped oscillation behavior before settling to a constant angle, whereas the ultrasonic system directly reaches a constant value.
Furthermore, although both start at zero degrees, their estimates show an initial offset of \reviseSec{\qty{0.2300}{\degree}} for the \ac{IMU} and an overshoot of \reviseSec{\qty{0.1500}{\degree}} in the ultrasonic angle. Once the motion stabilizes, they again converge to the initial angle of zero degrees.
In contrast, the ultrasound-based angle is derived from geometric measurements and therefore does not accumulate short-term drift, resulting in consistent convergence back to zero degrees.
A direct comparison between the ultrasonic system and the \ac{IMU} under dynamic conditions yields a \ac{MAE} of \reviseSec{\qty{0.2370}{\degree}}, which closely aligns with the data in the feasibility study in \cref{sec:feasibility}.

\subsection{System Performance}
This section investigates the system performance of the integrated ultrasonic ski edge angle detection system. While the primary focus of this work is to evaluate the feasibility of the proposed method, this section provides estimates of key parameters, such as latency and power consumption, to inform future embedded deployments.

For this evaluation, latency denotes the delay between acquiring a sensor sample and the corresponding angle estimate becoming available. It includes the time required for sensor data retrieval and subsequent processing. We measured an end-to-end latency of \qty{30.31}{\milli\second}. The dominant contributor is the 7-tap \ac{FIR} filter, which introduces a theoretical group delay of \qty{30}{\milli\second}.
In comparison, sensor data retrieval requires \qty{140}{\micro\second} for the \textit{TDK CH101} and \qty{61}{\micro\second} for the \textit{TDK ICM-20948}. The subsequent arithmetic processing adds \qty{140}{\micro\second} and \qty{64}{\micro\second}, respectively.

The system's power consumption was characterized using the \textit{Nordic Power Profiler Kit II}. Each subsystem was profiled individually, including the two CH101 \ac{US} sensors, the in-shoe \ac{IMU}, and the \ac{MCU} \revise{running at \qty{128}{\mega\hertz}} responsible for data acquisition and signal processing. The system power consumption was measured by emulating a battery at \qty{3.7}{\volt} and recording the resulting current draw over 10 seconds.
This measurement includes the full end-to-end power consumption, including conversion losses caused by the \ac{PMIC}. The resulting values are summarized in \cref{tab:power_profiling}.

\begin{table}[thb]
    \centering
    \renewcommand{\arraystretch}{1.2}
    \caption{Power consumption of the individual sub-systems and the full system at \qty{1.8}{\volt}.}
    \label{tab:power_profiling}
    \begin{tabularx}{\columnwidth}{@{}
        p{0.3\columnwidth}                     % System column
        p{0.4\columnwidth}                     % Component column
        p{0.3\columnwidth}@{}}                 % Power column
        \toprule
        \textbf{System} & \textbf{Component} & \textbf{Power [µW]} \\
        \midrule
         \revise{2x US Transducer}  & CH101   &  \revise{\qty{68.2}{\micro\watt}}  \\
        Accelerometer         & ICM-20948             & \revise{\qty{124.5}{\micro\watt}} \\ 
        \revise{Microcontroller} & \revise{nRF5340}   & \revise{\qty{1.092}{\milli\watt}} \\
        \midrule
        \textbf{Full System}    &           & \revise{\textbf{\qty{1.28}{\milli\watt}}}   \\ %674,76 uA
        \bottomrule
    \end{tabularx}
\end{table}

With an average power consumption of only \revise{\qty{1.28}{\milli\watt}}, the system's battery capacity can be significantly reduced, affecting the overall system weight.
A lightweight lithium-ion cell such as the \textit{Renata ICP341018PM}, which provides \qty{33}{\mAh} at only \qty{1.7}{\gram}, still offers sufficient energy to operate throughout a full training day, which typically comprises a three-hour morning session and a three-hour afternoon session. This results in an overall system weight of only \qty{18.6}{\gram}.

\section{Limitations \& Future Work}\label{sec:limitations}
The proposed system has been extensively tested and validated under laboratory conditions; however, transitioning to field deployment introduces new challenges. This section discusses potential factors that may impact sensing performance and the system's mechanical and electrical stability.

\subsection{Environmental Influences on Sensing}
Ski jumping is practiced year-round, with competitions typically held in winter and training sessions conducted during summer. These periods expose the system to varying environmental conditions that influence the sensor's performance.

One important factor to investigate is the wind speed and the turbulent airflow around the ski boots during jumps. Rapid fluctuations in air pressure and velocity can disrupt the propagation of ultrasonic waves, leading to acoustic distortion and inaccurate distance measurements. \reviseSec{Preliminary tests were conducted under controlled laboratory conditions and did not indicate measurable disturbances in the distance measurements. However, these conditions do not reflect the aerodynamic environment encountered in ski jumping, where velocities can exceed 90\;km\slash h~\cite{Liu2025}. While the underlying physical principles suggest that the short measurement range may limit susceptibility to airflow disturbances, the system has not yet been experimentally validated under such conditions. Therefore, comprehensive evaluation under realistic field conditions remains an important direction for future work.}

The different training seasons expose the system to varying environmental conditions.
As ultrasonic measurements are dependent on humidity and temperature~\cite{wells2020physics}, these factors must be considered to ensure reliable system performance under real-world conditions. The underlying physical relationships are well understood and can be corrected dynamically \cite{Tsai2005, Wong1986} using \cref{eq:velocity_correction}.

\revise{\begin{equation}
v_{\text{standard}} = \sqrt{\frac{T\,\gamma\,\mathrm{R}}{M}}
\label{eq:velocity_correction}
\end{equation}}

The temperature-dependent speed of sound described in \cref{eq:velocity_correction} depends on several constants: \revise{\( \mathrm{R} \)} is the universal gas constant, \(\gamma\) is the adiabatic index of air, \(M\) is the molar mass of air, and \(T\) is the absolute temperature in Kelvin.
To account for humidity, the empirical correction from \cref{eq:humidity_correction} can be applied.

\begin{equation}
    v_{\text{corrected}}(T,RH) = v_{\text{standard}} + (0.61T_{\text{Celsius}}) + (0.012RH)
    \label{eq:humidity_correction}
\end{equation}

In this expression,
\(T_{Celsius}\) denotes the temperature in degrees Celsius, and \(RH\) represents the relative humidity in percent \cite{Tsai2005, PANDA2016574}.

Other environmental factors, such as fog, rainfall, and snowfall, may also affect system performance.
While these effects have not been thoroughly studied in ski jumping, related work on sonar sensors in mobile robots and autonomous vehicles exists \cite{5509804, 8666747}.
Given the short operating range, significant signal degradation is unlikely but should be verified through field testing.

\subsection{Impact on System Integration}
When deploying the system for real-world field testing, its integration must be tailored to the specific application conditions. In ski jumping, the system is exposed to adverse environmental factors such as wind, water, fog, and dust. Therefore, a waterproof design is essential.
This can be achieved by embedding the sensors and electronics into the ski boot, thereby improving robustness against turbulence and enhancing overall measurement reliability.

\section{Conclusion}\label{sec:conclusion}
In professional ski jumping, the ski edge angle is a critical performance parameter because it directly influences aerodynamic lift. Prior work indicates that maintaining an edge angle between \ang{0} and \ang{5} improves the lift-to-drag ratio. However, precise control of the edging angle requires accurate and timely feedback.

This work presented the theoretical analysis, system integration, and experimental evaluation of a multi-sensor system for ski edge angle estimation in ski jumping. The system integrates two ultrasonic \ac{ToF} sensors and an \ac{IMU} into a single unit. Unlike existing approaches that rely on multiple devices, including ski-mounted components, the proposed design reduces system mass, simplifies handling, and improves overall integration.

Laboratory experiments showed that the system achieved an angular resolution of \reviseSec{\ang{0.4500}} and a \ac{MAE} of \reviseSec{\ang{0.2640}}. The coefficient of determination exceeded 99\%. A comparison of the two sensing modalities demonstrated that the ultrasonic measurements are robust against drift and bias because the angle is derived directly from geometric relationships. The measured end-to-end system latency of \qty{30.31}{\milli\second} is compatible with real-time feedback applications.

With a power consumption of \revise{\qty{1.28}{\milli\watt}} and a total system mass of \qty{18.6}{\gram}, the results demonstrate the feasibility of integrating ultrasonic angle estimation into a ski boot in a lightweight and unobtrusive form factor. The presented system provides a foundation for future development of ultrasonic wearable solutions for ski edge angle estimation and boot inclination sensing in ski jumping.

\section*{ACKNOWLEDGMENT}
\revise{The authors would like to acknowledge SwissSki for providing the binding system used in the experimental recordings with the integrated setup. The authors further thank Alexander Pointner for his support during the initial phase of the project and for facilitating valuable contacts and exchanges.}

\bibliographystyle{IEEEtran}
\bibliography{
    bib/references,
    bib/us_skiboot}

\begin{IEEEbiography}[{\includegraphics[width=1in,height=1.25in,clip,keepaspectratio]{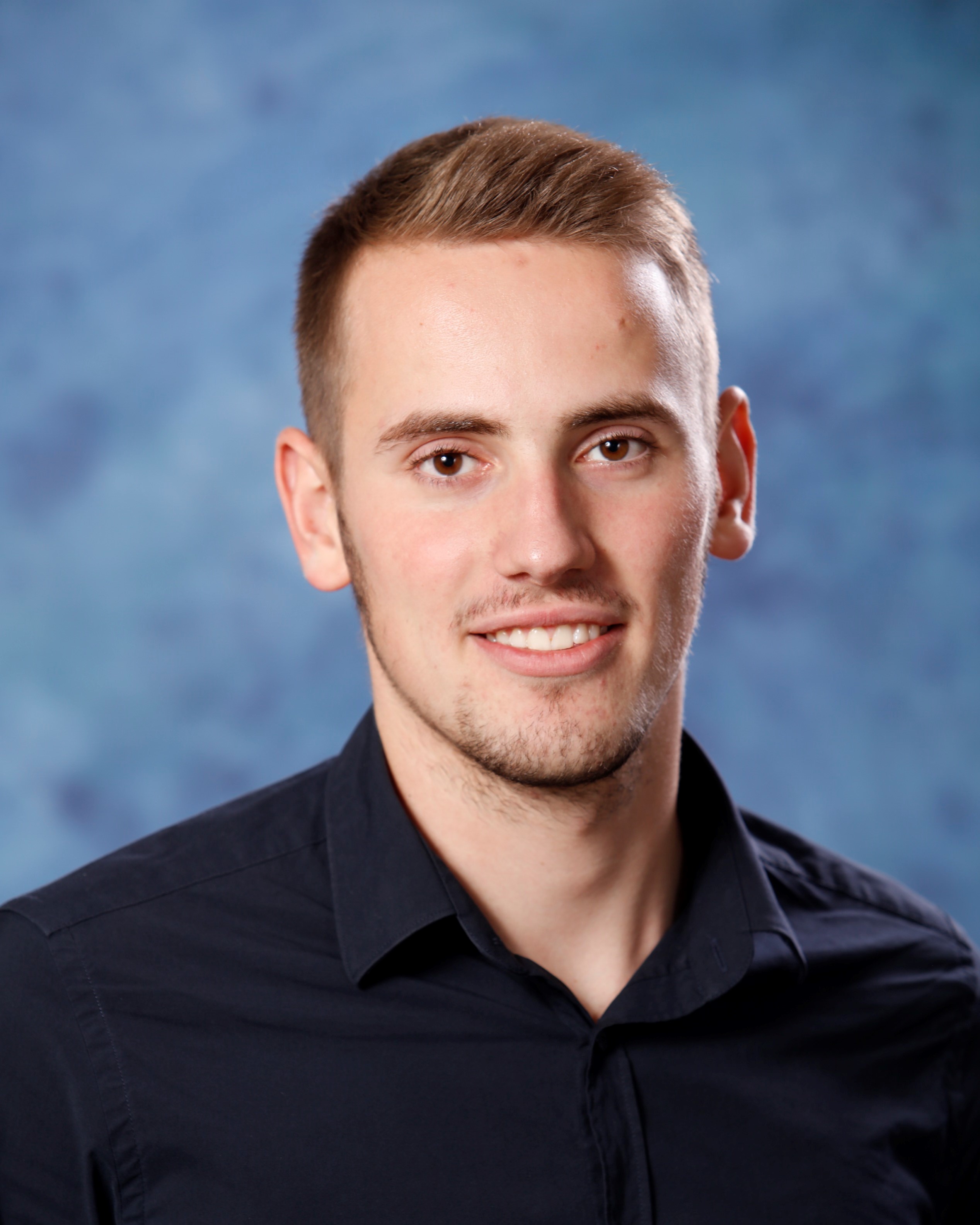}}]{Ivan Simeonov} received a B.Sc. in biomedical technology from CTU Prague in 2023. He is currently pursuing a M.Sc. degree in biomedical engineering at ETH Zurich with a specialization in bioelectronics. His current research interests include medical and biomedical device development, with a focus on photoplethysmography, capnography, gas flow, and POC wearables.
\end{IEEEbiography}

\begin{IEEEbiography}[{\includegraphics[width=1in,height=1.25in,clip,keepaspectratio]{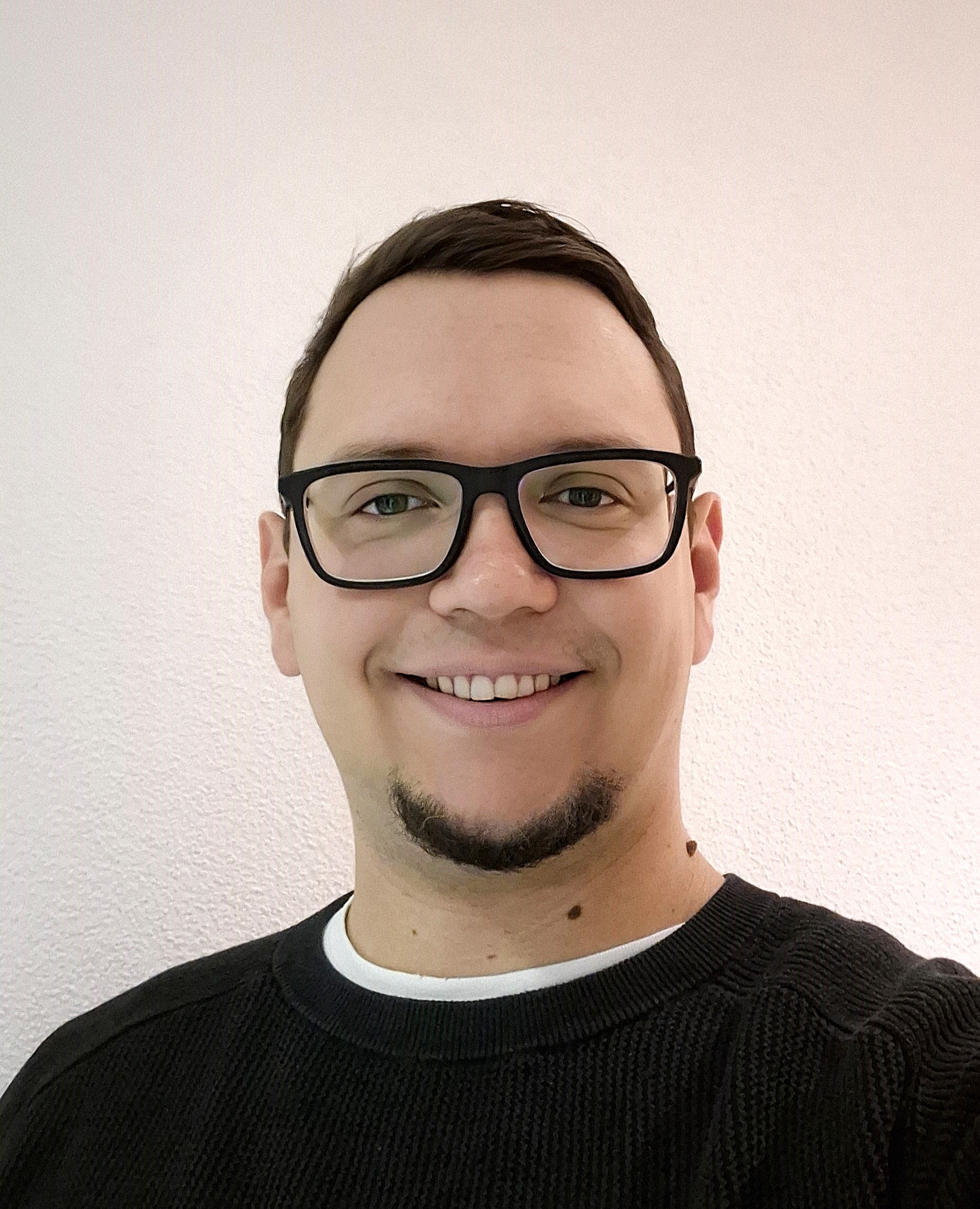}}]{Lukas Schulthess} (Graduate Student Member, IEEE) received the Swiss Certificate of Competence (EFZ) as an electronic technician in 2014. He received the B.Sc. and the M.Sc. degrees in electronics engineering and information technology from ETH Zurich (Zurich, Switzerland), in 2020 and 2021, respectively. He is currently pursuing a Ph.D. degree at ETH Zurich (Zurich, Switzerland). His research interests include ultra-low power and miniaturized self-sustainable sensor nodes, wireless body area networks, and energy harvesting.
\end{IEEEbiography}

\begin{IEEEbiography}[{\includegraphics[width=1in,height=1.25in,clip,keepaspectratio]{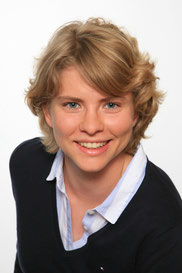}}]{Hanna Mueller}
(Graduate Student Member, IEEE) received the B.Sc. and M.Sc. and Ph.D. degrees in electrical engineering and information technologies from ETH Zurich (Zurich, Switzerland), in 2017, 2020, and 2025 respectively. Her Ph.D. degree with the Integrated Systems Laboratory focused on low-power systems, wireless sensor networks and onboard intelligence - especially for obstacle avoidance and localization of nano-drones.\end{IEEEbiography}

\begin{IEEEbiography}[{\includegraphics[width=1in,height=1.25in,clip,keepaspectratio]{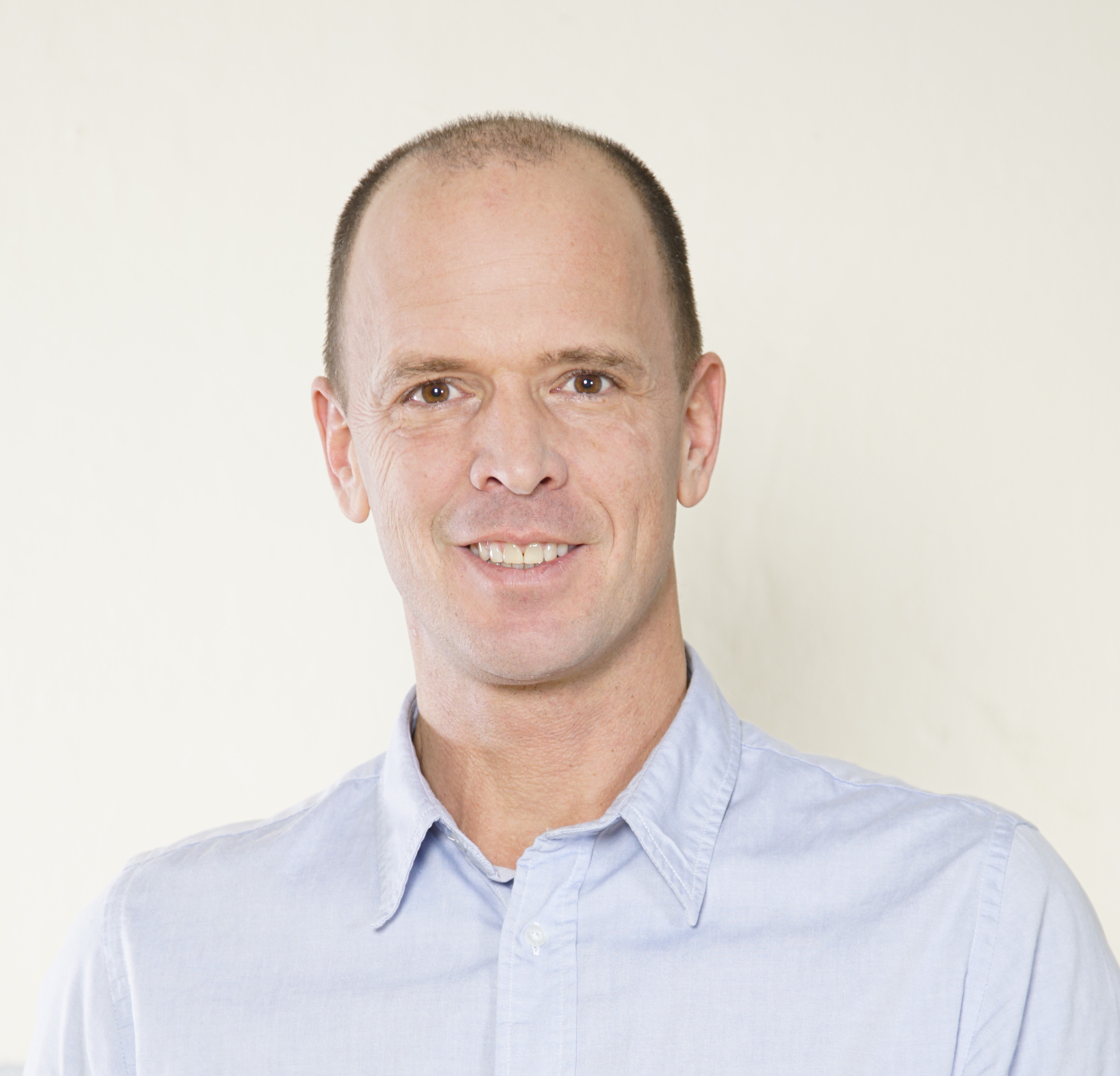}}]{Marc Nölke} is a certified trainer with the German Olympic Sports Confederation (DOSB) and a former competitive ski jumper. He has worked as a trainer and performance specialist for national ski jumping teams in Austria, Germany, Poland, Finland, and the USA. His work combines practical high-performance training and rehabilitation experience with a mechanistic analysis of biomechanics, sensor technology, and neurophysiological movement control. His professional focus is on the development of test/retestable feedback and training systems for performance-oriented and clinical applications.
\end{IEEEbiography}

\begin{IEEEbiography}[{\includegraphics[width=1in,height=1.25in, clip,keepaspectratio]{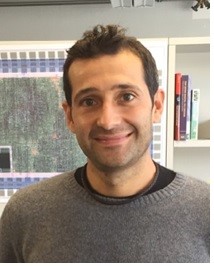}}]{Michele Magno} (Fellow, IEEE) is currently a Senior Scientist at ETH Zurich (Zurich, Switzerland), in the Department of Information Technology and Electrical Engineering. Since 2020, he is leading the D-ITET center for project-based learning at ETH. He received his master's and Ph.D. degrees in electronic engineering from the University of Bologna, Italy, in 2004 and 2010, respectively. He has been working at ETH since 2013 and has become a visiting lecturer or professor at the University of Nice Sophia, Enssat Lannion, University of Bologna, and Mid Sweden University. His current research interests include smart sensing, low-power machine learning, wireless sensor networks, wearable devices, energy harvesting, and low-power management techniques. He has authored over 280 publications in international journals and conferences, earning best paper awards at IEEE conferences and recognition for industrial projects and patents.
\end{IEEEbiography}

\begin{IEEEbiography}[{\includegraphics[width=1in,height=1.25in,clip,keepaspectratio]{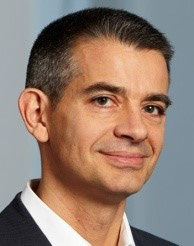}}]{Luca Benini} (Fellow, IEEE) received the Ph.D. degree from Stanford University (Stanford, USA). He holds the Chair of digital circuits and systems at the ETH Zurich (Zurich, Switzerland). He is a Full Professor at the University of Bologna (Bologna, Italy). His research interests include energy-efficient parallel computing systems, smart sensing micro-systems, and machine learning hardware. He is a Fellow of the ACM and a Member of the Academia Europaea. He is the recipient of the 2016 IEEE CAS Mac Van Valkenburg Award, the 2020 EDAA Achievement Award, and the 2020 ACM/IEEE A. Richard Newton Award, and the 2023 IEEE CS E.J. McCluskey Award.
\end{IEEEbiography}

\begin{IEEEbiography}[{\includegraphics[width=1in,height=1.25in,clip,keepaspectratio]{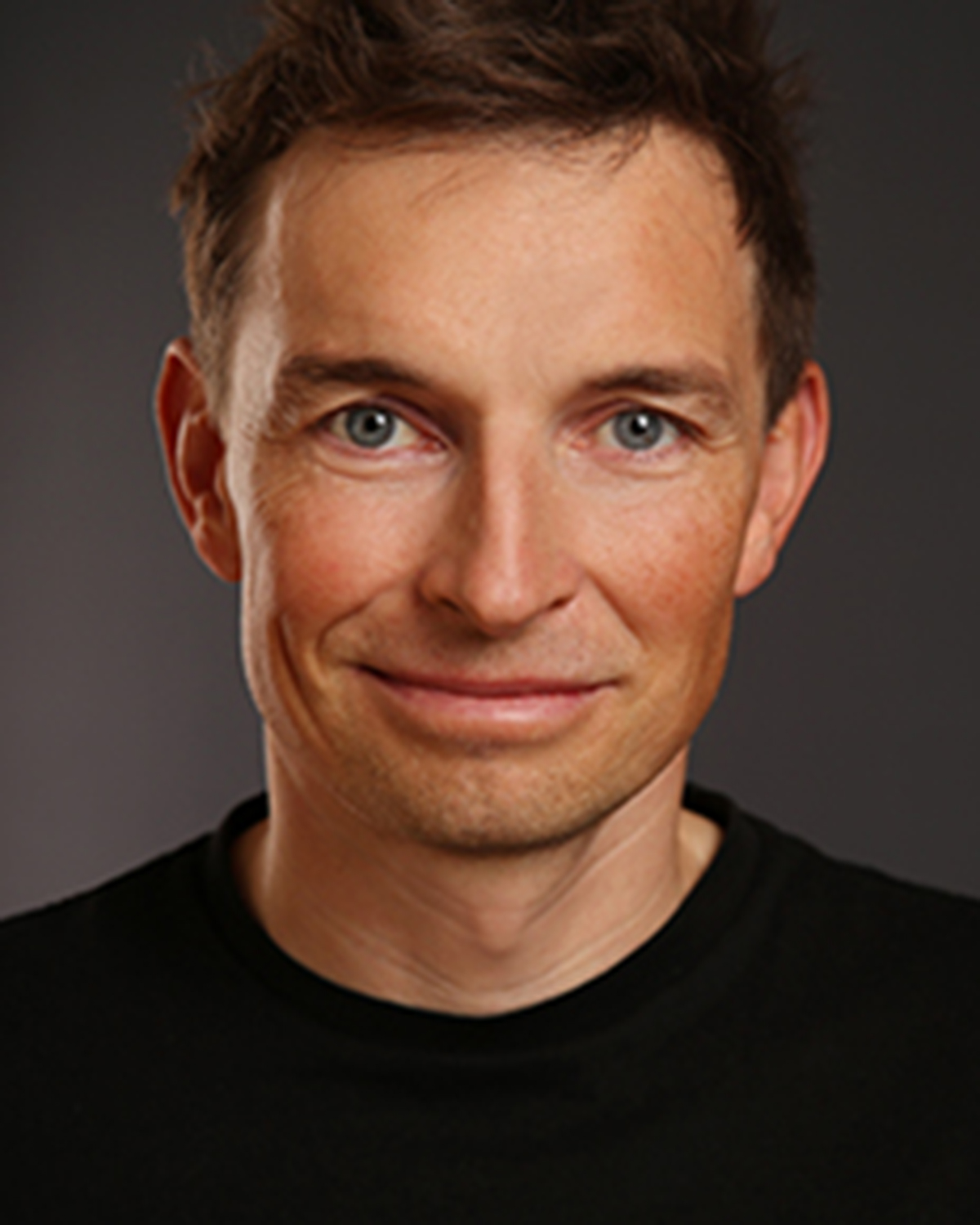}}]{Christoph Leitner} (Senior Member, IEEE) is a Principal Investigator under the SNSF Ambizione program at the Integrated Systems Laboratory in the Department of Information Technology and Electrical Engineering at ETH Zurich (Zurich, Switzerland) since 2026, where he conducted also his postdoctoral research under Prof. Luca Benini from 2023 to 2025. He received the Ph.D. degree in Biomedical Engineering in 2022 and the M.Sc. degree in Mechanical Engineering and Economics in 2006, both from Graz University of Technology (Graz, Austria). The focus of his research is on human-centred sensing and cyber-physical systems, with a particular emphasis on acoustics, signal processing and high-frequency mixed-signal electronics for wearable and distributed sensing and computing platforms.
\end{IEEEbiography}

\end{document}